\documentclass[preprint]{aastex}
\begin{document}
 
  \newcommand{\Cerenkov}{\v Cerenkov } 
  \newcommand{\Check}{{\Large \emph{\textbf{CHECK THIS}}}}
  \newcommand{\Flux}{$1.6 \times 10^{-8} \pm 0.5\times10^{-8}
    (\mathrm{stat}) \pm 0.3 \times 10^{-8}
    (\mathrm{sys})~\mathrm{photons~m^{-2}~s^{-1}}$}

  \title{ TeV Gamma-Ray Observations of the Galactic Center }
  
  \author{ K. Kosack,\altaffilmark{1}
    H. M. Badran, \altaffilmark{2}
    I. H. Bond,\altaffilmark{3}
    P. J. Boyle,\altaffilmark{4}
    S. M. Bradbury,\altaffilmark{3}
    J. H. Buckley,\altaffilmark{1}
    D. A. Carter-Lewis,\altaffilmark{6}
    M. Catanese,\altaffilmark{5}
    O. Celik,\altaffilmark{7}
    V. Connaughton,\altaffilmark{21}
    W. Cui,\altaffilmark{8}
    M. Daniel,\altaffilmark{6}
    M. D'Vali,\altaffilmark{3}
    I. de la Calle Perez,\altaffilmark{3}
    C. Duke,\altaffilmark{9}
    A. Falcone,\altaffilmark{8}
    D. J. Fegan,\altaffilmark{10}
    S. J. Fegan,\altaffilmark{5}
    J. P. Finley,\altaffilmark{8}
    L. F. Fortson,\altaffilmark{19}\altaffilmark{20}
    J. A. Gaidos,\altaffilmark{8}
    S. Gammell,\altaffilmark{10}
    K. Gibbs,\altaffilmark{5}
    G. H. Gillanders,\altaffilmark{11}
    J. Grube,\altaffilmark{3}
    J. Hall,\altaffilmark{12}
    T. A. Hall,\altaffilmark{13}
    D. Hanna,\altaffilmark{14}
    A. M. Hillas,\altaffilmark{3}
    J. Holder,\altaffilmark{3}
    D. Horan,\altaffilmark{5}
    A. Jarvis,\altaffilmark{7}
    M. Jordan,\altaffilmark{1}
    G. E. Kenny,\altaffilmark{11}
    M. Kertzman,\altaffilmark{15}
    D. Kieda,\altaffilmark{12}
    J. Kildea,\altaffilmark{14}
    J. Knapp,\altaffilmark{3}
    H. Krawczynski,\altaffilmark{1}
    F. Krennrich,\altaffilmark{6}
    M. J. Lang,\altaffilmark{11}
    S. Le Bohec,\altaffilmark{6}
    R. W. Lessard,\altaffilmark{8}
    E. Linton,\altaffilmark{4}
    J. Lloyd-Evans,\altaffilmark{3}
    A. Milovanovic,\altaffilmark{3}
    J. McEnery, \altaffilmark{17}
    P. Moriarty,\altaffilmark{16}
    D. Muller,\altaffilmark{4}
    T. Nagai,\altaffilmark{12}
    S. Nolan,\altaffilmark{8}
    R. A. Ong,\altaffilmark{7}
    R. Pallassini,\altaffilmark{3}
    D. Petry,\altaffilmark{17}
    B. Power-Mooney,\altaffilmark{10}
    J. Quinn,\altaffilmark{10}
    M. Quinn,\altaffilmark{16}
    K. Ragan,\altaffilmark{14}
    P. Rebillot,\altaffilmark{1}
    P. T. Reynolds,\altaffilmark{18}
    H. J. Rose,\altaffilmark{3}
    M. Schroedter,\altaffilmark{5}
    G. H. Sembroski,\altaffilmark{8}
    S. P. Swordy,\altaffilmark{4}
    A. Syson,\altaffilmark{3}
    V. V. Vassiliev,\altaffilmark{7}
    S. P. Wakely,\altaffilmark{4}
    G. Walker,\altaffilmark{12}
    T. C. Weekes,\altaffilmark{5}
    J. Zweerink,\altaffilmark{7}}

  \altaffiltext{1}{Department of Physics, Washington University, St. Louis, MO 63130, USA}  
  \altaffiltext{2}{Physics Department, Tanta University, Tanta, Egypt}
  \altaffiltext{3}{Department of Physics, University of Leeds, Leeds, LS2 9JT, Yorkshire, England, UK}
  \altaffiltext{4}{Enrico Fermi Institute, University of Chicago, Chicago, IL 60637, USA}
  \altaffiltext{5}{Fred Lawrence Whipple Observatory, Harvard-Smithsonian CfA, P.O. Box 97, Amado, AZ 85645-0097} 
  \altaffiltext{6}{Department of Physics and Astronomy, Iowa State University, Ames, IA 50011-3160, USA}
  \altaffiltext{7}{Department of Physics, University of California, Los Angeles, CA 90095-1562, USA}
  \altaffiltext{8}{Department of Physics, Purdue University, West Lafayette, IN 47907, USA}
  \altaffiltext{9}{Department of Physics, Grinnell College, Grinnell, IA 50112-1690, USA}
  \altaffiltext{10}{Experimental Physics Department, National University of Ireland, Belfield, Dublin 4, Ireland}
  \altaffiltext{11}{Department of Physics, National University of Ireland, Galway, Ireland}
  \altaffiltext{12}{High Energy Astrophysics Institute, University of Utah, Salt Lake City, UT 84112, USA}
  \altaffiltext{13}{Department of Physics and Astronomy, University of Arkansas at Little Rock, Little Rock, AR 72204-1099, USA}
  \altaffiltext{14}{Physics Department, McGill University, Montre$\acute{a}$l, QC\,H3A\,2T8, Canada}
  \altaffiltext{15}{Department of Physics and Astronomy, DePauw University, Greencastle, IN 46135-0037, USA}
  \altaffiltext{16}{School of Science, Galway-Mayo Institute of Technology, Galway, Ireland}
  \altaffiltext{17}{University of Maryland, Baltimore County and NASA/GSFC, USA}
  \altaffiltext{18}{Department of Applied Physics and Instrumentation, Cork Institute of Technology, Cork, Ireland}
  \altaffiltext{19}{Department of Astronomy and Astrophysics,
  University of Chicago, Chicago, IL, USA}
  \altaffiltext{20}{Astronomy Department, Adler Planetarium and
  Astronomy Museum, Chicago, Il, USA.}
  \altaffiltext{21}{Gamma-Ray Astrophysics Group National Space
  Science and Technology Center Huntsville, Alabama}
    
  \email{kosack@hbar.wustl.edu\ \ buckley@wuphys.wustl.edu}
  
  \begin{abstract}
    We report a possible detection of TeV gamma-rays from the Galactic
    Center by the Whipple 10m gamma-ray telescope.  Twenty-six hours
    of data were taken over an extended period from 1995 through 2003
    resulting in a total significance of 3.7 standard deviations.  The
    measured excess corresponds to an integral flux of \Flux
    above an energy of $2.8\ \mathrm{TeV}$, roughly 40\% of the flux
    from the Crab Nebula at this energy.  The 95\% confidence region
    has an angular extent of about 15 arcmin and includes the position
    of Sgr A*. The detection is consistent with a point source and
    shows no evidence for variability. 

  \end{abstract}
  
  \maketitle

  \section{Introduction}

  The central region of our galaxy is now thought to contain a
  super-massive black-hole of 2.6$\times 10^6 M_\odot$
  \citep{2002AAS...201.6804G,2002Natur.419..694S} coincident with the
  unresolved radio source Sgr A* \citep{balick1974}. Chandra
  observations reveal X-ray emission from an unresolved point source
  as well as an extended structure ($\sim 1.5 \ \mathrm{arcsec}$),
  both of which appear to be physically associated with Sgr A* (e.g.,
  \citet{2003ApJ...591..891B}).  The recent discovery of hour-scale
  X-ray \citep{2001Natur.413...45B} and rapid IR flaring
  \citep{2004ApJ...601L.159G} point to an active nucleus, albeit with
  very low bolometric luminosity compared with the luminosity inferred
  from the Bondi accretion rate or with that which is typical of more
  powerful AGNs.  More recently, \emph{INTEGRAL} (\emph{the
  International Gamma-Ray Astrophysics Laboratory}) has detected
  time-variable 20-100 keV emission from within 0.$^{'}$9 of Sgr
  A*\citep{2004ApJ...601L.163B}.  Polarization measurements show the
  signature of synchrotron radiation in a Keplarian accretion disk
  \citep{2002ApJ...566L..77L}. Taken together, these multi-wavelength
  data are not easily described by a one-component model, and the
  current theoretical framework combines thermal emission from a
  radiatively inefficient Keplarian accretion flow with synchrotron
  inverse-Compton emission produced either by electrons accelerated in
  the disk or further out in a hypothetical jet-like outflow
  (e.g. \citet{2002ApJ...566L..77L,2002A&A...383..854Y,2003astro.ph..4125Y}).
  From the present measurements, the maximum energy of the non-thermal
  electron distribution in the jet models is ambiguous and theories
  alternately explain the high-energy emission as inverse-Compton or
  the high-energy extension of the synchrotron spectrum; gamma-ray
  measurements may eventually break this degeneracy.

  The EGRET experiment detected a strong unidentified source of GeV
  gamma-rays marginally consistent with the position of the Galactic
  Center \citep{1999ApJS..123...79H}.  Both the Whipple and Cangaroo
  groups have presented preliminary evidence for TeV emission at the
  position of Sgr A* as well \citep{buckley97, cangaroo2003, 
  kosack2003}.  \citet{dingus2002} re-analyzed the higher energy
  gamma-ray data from EGRET and found that the most likely position of
  the EGRET source may be offset from Sgr A*.  However, systematic
  uncertainties in the gamma-ray background models and limited angular
  resolution make the analysis of the source in the Galactic Center
  region difficult. Observations of the Galactic Center are
  complicated since Sgr A* is surrounded by a dense cluster of stars
  and stellar remnants (including low-mass X-ray binaries and black
  hole candidates), molecular clouds, and a large structure that may
  be the remnant of a powerful supernova remnant, Sgr A East
  \citep{2003ApJ...596.1035F}.  Source confusion is particularly
  difficult for high-energy gamma-ray observations given the limited
  angular resolution of present experiments.


  High energy gamma-ray observations of the Galactic Center are also
  the subject of particular theoretical interest given the possibility
  of detecting halo dark matter in our galaxy
  (e.g. \citet{1998APh.....9..137B}).  Sgr A*, at the dynamical center
  of our galaxy, may well be surrounded by a cusp or spike in the dark
  matter halo distribution (e.g.  \citet{1991ApJ...378..496D,
  1996ApJ...462..563N,1999PhRvL..83.1719G, 2003astro.ph.11594M}).
  Annihilation of these hypothetical weakly interacting massive
  particles could also contribute to the luminosity in the vicinity of
  Sgr A* in the radio through gamma-ray waveband.  Annihilation of
  dark matter would be enhanced by a factor proportional to the
  density squared, and might result in an observable gamma-ray line
  (from direct annihilation to gamma rays) as well as continuum
  emission (from secondary products of annihilation to quarks and
  fermions) \citep{1987ApJ...313L..47S, 1989pafe.conf..255B,
  1989PhRvD..39.3549R,
  1989PhRvD..40.2549G,1989ApJ...343..169S,1995PhRvD..51.3121J}. The
  presence of a massive black hole could further steepen the density
  profile of the dark matter halo, producing very high radio and
  gamma-ray fluxes that exceed the observational upper bounds
  \citep{1999PhRvL..83.1719G}. The details of the halo model on scales
  $<100\mathrm{pc}$ and the formation history of the central black
  hole are critical to predicting the gamma-ray flux, but are,
  unfortunately, still poorly understood.

  Given the limited angular resolution of GeV and TeV instruments, a
  number of different sources could contribute to a signal near
  the Galactic Center. The key to distinguishing between all of the
  possible emission scenarios is to measure the position, angular
  extent, variability and spectrum of the gamma-ray signal. Here we
  present first results from an analysis of Whipple telescope data. In
  \S\ref{sec:method} we describe the observational method and data
  analysis procedure used to observe the Galactic Center at TeV
  energies. In \S\ref{sec:analysis} and \S\ref{sec:results}, we
  discuss a possible weak detection and consider its impact on various
  gamma-ray production scenarios in \S\ref{sec:discussion}.

  \section{Method}
  \label{sec:method}

  \begin{table}
    \begin{center}
    \begin{tabular}[t]{c c c c}
      \hline\hline
      Season(s) & F.O.V ($^\circ$)& Pixels & Pixel Diameter($^\circ$) \\
      \hline
      1995-1996 & 3.1  & 109 & 0.26 \\
      1996-1997 & 3.1 & 151 & 0.26 \\
      1999-2003 & 2.4 & 379 & 0.12\\
      \hline\hline
    \end{tabular}
    \end{center}
    \caption{\label{tab:cameras} Camera specifications during seasons
    where Galactic Center data were taken.}
  \end{table}

  Imaging Atmospheric \Cerenkov Telescopes (IACTs), like the Whipple
  Gamma-ray Observatory, detect high energy photons by imaging the
  flashes of \Cerenkov light emitted by secondary particles in
  gamma-ray induced air showers. The Whipple Telescope's 10m mirror
  focuses the faint UV/blue \Cerenkov flashes into a camera consisting
  of 379 photomultiplier tube pixels (for the latest seasons---see
  Table \ref{tab:cameras} for previous seasons).  Off-line software
  analysis characterizes each candidate shower image, separates signal
  (gamma-ray-like) from background (cosmic-ray-like) events, and
  determines the point of origin and energy of each gamma ray.

  Whipple gamma-ray data are traditionally taken as a series of 28
  minute exposures, each of which is followed by an off-source run
  which is offset 30 minutes in right ascension for background
  subtraction. In the case of Sgr A*, data were taken off-source before
  the on-source observations due to a bright star field in the region
  30 minutes past the Galactic Center's position.  The analysis method
  used here (see \S\ref{sec:analysis}) was modified to give sensitivity at
  large zenith angles and to provide a two-dimensional map of the TeV
  emission in a three-degree diameter field-of-view surrounding the
  Galactic Center.

  \section{Data Analysis}
  \label{sec:analysis}

  The Galactic Center ($\alpha=17^{h}45^{m}40^{s},\
  \delta=-29^{\circ}00'28''$, J2000) transits at a very large zenith
  angle ($61^\circ$) as seen from the Whipple Observatory
  ($31.7^\circ$ N latitude) which significantly alters the shower
  geometry and threshold energy.  To properly account for the effects
  of LZA observations, special techniques that go beyond the standard
  Whipple analysis were required.  Furthermore, the brightness of the
  galactic plane near the Galactic Center results in an increase in
  energy threshold and, if not compensated for, a systematic bias in
  the observed excess. Pedestal events, containing no image, are
  injected at random intervals throughout the run for calibration of
  Poisson fluctuations in the night sky background. Both on and
  off-source data are analyzed in the same manner, and Gaussian
  deviates are added to the pixel signals to bring the background
  noise up to the same level in both runs. After this procedure, only
  pixels with signals well above the noise level are included in
  further image processing. This Gaussian \emph{padding} combined with
  a high software trigger threshold largely removes systematic
  biases arising from brightness differences, but increases the energy
  threshold.

  Using the techniques based on the moment fitting procedure outlined
  in \citet{1993ApJ...404..206R}, we parameterize the roughly
  elliptical gamma-ray images by calculating moments of the light
  distribution in the camera.  Geometric selection criteria based on
  these parameters allow for the rejection of background
  (e.g. cosmic-ray induced showers). The first moments give the
  centroid of the image, the second moments give the WIDTH (minor
  axis) and LENGTH (major axis) and orientation angle of the image.
  The elongation of the ellipse is used to determine the point of
  origin of each shower image using the formula $ \delta = \epsilon
  \left( 1-\frac{\mathrm{WIDTH}}{\mathrm{LENGTH}} \right)$, where
  $\delta$ is the displacement of the point of origin from the image
  centroid and $\epsilon$ is the elongation factor (determined by
  simulations).  Of the two possible points of origin for each image,
  the asymmetry (or skew) of the shower is used to select the correct
  one whenever possible. This procedure is similar to that described
  in \citet{2001APh....15....1L}.

  To determine the pointing error in the telescope, we look at the
  pedestal variation for each tube.  The presence of visible light
  from a star or other source of sky brightness in the field of view
  adds to the pedestal variance in the corresponding photomultiplier
  tube.  Using this effect, we can generate a crude optical image of
  the sky by de-rotating the camera to a common orientation (since the
  field of view rotates with time for an altitude-azimuth
  telescope) and accumulating the pedestal variations of each pixel
  into a two-dimensional histogram.  Using this technique, an optical
  \emph{sky-brightness} map is generated for each observation. By
  comparing the bright spots in the Sgr A* image (or special runs
  where the telescope is pointed at a nearby \emph{pointing-check}
  star) we can obtain an absolute measure of the pointing error.  In
  addition, we cross-correlate each pair of maps to determine the
  relative pointing offset between them.  We only keep runs that have
  the correct star field and have a relative pointing offset (compared
  with the other Sgr A* runs) which is less than the diameter of one
  pixel. This lessens the possibility of accidentally including an
  observation with large pointing errors or which was mislabeled.
  Using this method, we find residual pointing errors of $\pm
  0.1^\circ$ and an absolute offset of $0.14^\circ$. These errors are
  attributed in part to the fact that observations have been made near
  the balance point of the telescope and near the horizontal position
  where flexure of the optical support structure is maximal.

  Sgr A* transits at roughly $30^\circ$ elevation as observed from
  the latitude of the Whipple Observatory.  Large zenith angle (LZA)
  observations require several modifications to the standard analysis
  due to changes in shower geometry.  In addition, we have data on Sgr
  A* that spans several epochs during which the Whipple camera was
  upgraded twice (see Table \ref{tab:cameras}).  In order to combine
  all of these data, we needed our gamma-ray selection criteria to
  scale properly with changes in zenith angle, the camera throughput
  factor (i.e. the ratio of photons hitting the mirror to digital
  counts in the digitized PMT signals), and the pixel
  size. In addition, for subsequent spectral analysis, we designed our
  gamma-ray selection procedure to scale with energy to minimize
  spectral biases.

  To achieve this goal, we started with the standard Whipple Telescope
  data analysis procedure \citep{2001APh....15....1L} and developed a new set
  of gamma-ray selection criteria that scale with zenith angle and
  energy according to a semi-empirical form derived from simulations
  and optimization of LZA Crab Nebula data.  Our criteria also
  incorporate the geometric effects of differing cameras, allowing us
  easily to combine all of the data present for the Galactic Center.
  The Crab Nebula was used for optimization and calibration
  because it is a bright, steady gamma-ray source with a known
  spectrum. In addition, an independent set of Crab Nebula
  observations taken at LZA were used to verify that the selection
  criteria were more efficient than the standard analysis procedure.
  
  To correct for changes in the overall light sensitivity (throughput) of the
  camera between epochs, we look at muon events in data taken at each
  epoch. Muons, which show up as bright arcs of \Cerenkov light in the
  camera, are useful for throughput calibration because the light per
  unit arc-length from a muon event is nearly constant regardless
  of the impact parameter and angle of the trajectory.  A measure of
  the throughput of the telescope can be found by making a histogram
  of the \emph{signal/arc-length} distribution of a large set of muon
  events. To get an absolute calibration, this distribution for real
  observations was compared with a set of simulated muon events which
  were generated using the Grinnell/ISU (GrISU) simulation package to produce a
  large number of H and He showers that in turn produce muons as
  secondary particles.  For instance, the relative throughput was
  found to have changed by a factor of 2.22 between 1995 and 2001. We
  scale the software trigger threshold and energy estimator by this
  factor so the trigger cuts are consistent across observations. This
  method also serves to calibrate the simulations used to determine
  the peak energy at LZA (see \S\ref{sec:results}).

  As one observes at increasing zenith angles, the distance to 
  the core of the air-shower increases and thus the angular size of the shower
  and parallactic displacement of the image centroid are reduced.  To
  derive the scaling laws, we first assume that the WIDTH and LENGTH
  of gamma-ray air shower images is approximately proportional to
  $\cos^\alpha\theta$, where $\theta$ is the
  zenith angle and $\alpha$ is a constant. Additionally, air-shower
  simulations show that LENGTH and WIDTH also scale as the logarithm
  of the energy, which is proportional to the SIZE (total camera
  signal) of the event.

\begin{figure}[ht!]
    \includegraphics[width=\columnwidth]{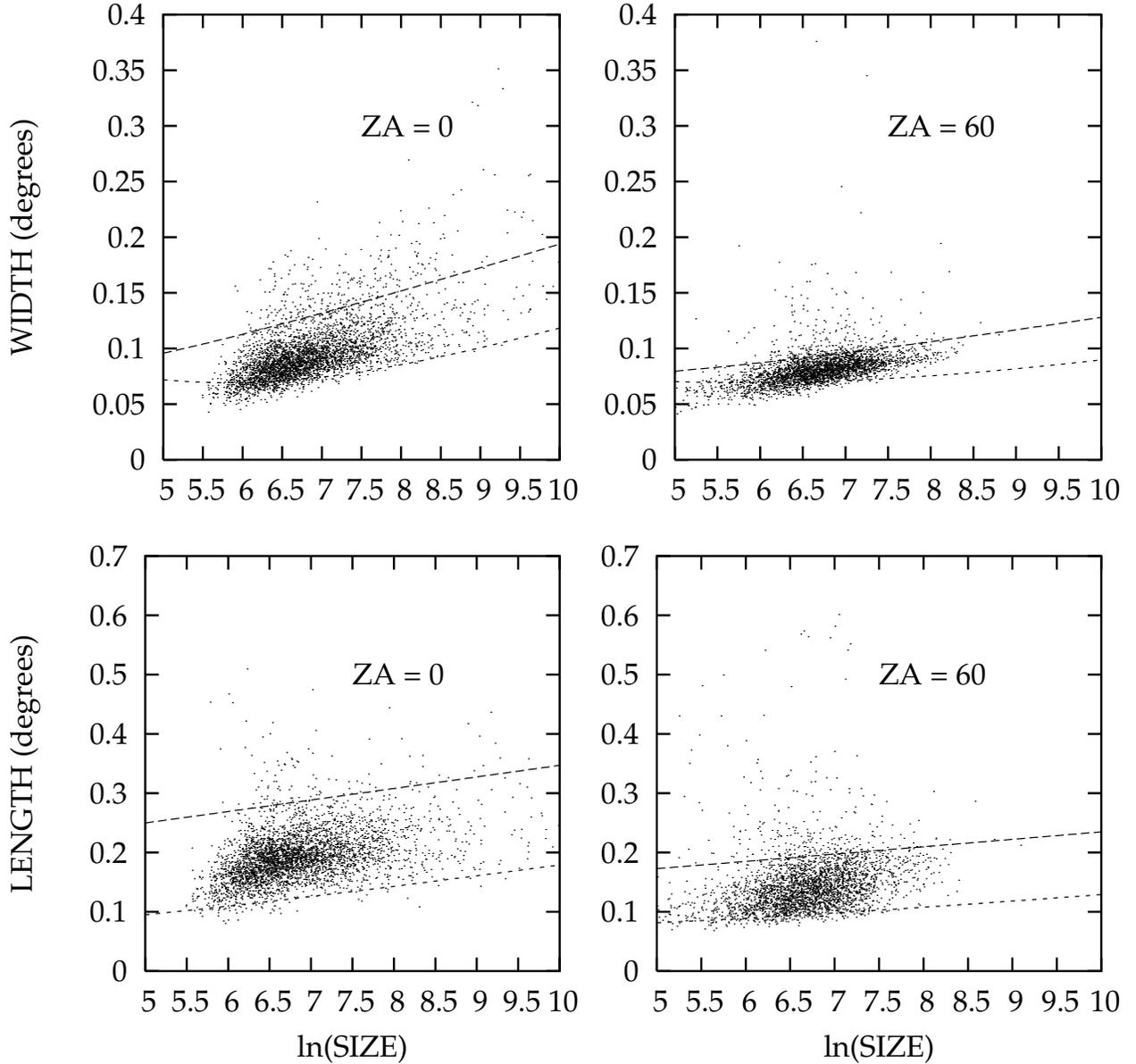}
    \caption{ \label{fig:zcuts} Plots of our selection criteria for
    gamma-ray-like events for two of the parameters (LENGTH and
    WIDTH) for two different zenith angles ($0^\circ$ and $60^\circ$).
    The dotted lines show the upper and lower cuts on the respective
    parameters as a function of the SIZE (total signal) parameter. The
    dots are from simulations of gamma rays with a range of energies. }
  \end{figure}

  Combining these results, and removing the effects of the finite
  pixel size of the camera ($\sigma_{\mathrm{pix}}$) and the point
  spread function of the telescope ($\sigma_{\mathrm{psf}}$), the
  measured LENGTH ($L$), WIDTH ($W$), and the distance to the image
  centroid ($D$) can be converted to scaled values $L'$, $W'$ and $D'$
  by the following equations:
  \begin{eqnarray}
    \Sigma \equiv \ln(\mathrm{SIZE}) - 8.0 \nonumber \\
    \mathrm{L}' \simeq 
    \left[ \frac{ \mathrm{L}^2 
	- \sigma^2_{\mathrm{pix}} 
	- \sigma^2_{\mathrm{psf}}
    }{\cos^{1.5}\theta}
      \right]^\frac{1}{2} - 0.023\Sigma \\
    \mathrm{W}' \simeq
    \left[ \frac{
      \mathrm{W}^2 
      - \sigma^2_{\mathrm{pix}}
      - \sigma^2_{\mathrm{psf}}
    }{\cos^{1.2}\theta}
      \right]^\frac{1}{2} - 0.020\Sigma \\
    \mathrm{D}' \simeq 
    \frac{\mathrm{D}}{\cos\theta} 
  \end{eqnarray}
  The constant factors and cosine powers were derived from simulation
  fits and by optimization on Crab Nebula data (taken at a range of
  zenith angles) and then compared with simulations for calibration. Data
  selection criteria $0.125^\circ < L'< 0.3^\circ$,
  $0.05^\circ<W'<0.135^\circ$, $0.28^\circ<D'<2.2^\circ$ are applied
  to select candidate gamma-ray events (examples of these cuts are
  shown graphically in Figure \ref{fig:zcuts}). Cuts based on these
  intrinsic parameters were verified to be independent of zenith angle
  and camera design by application of this method to independent Crab
  Nebula data taken over the period 1994-2003.

  \section{Results}
  \label{sec:results}

  \begin{figure}[ht!]
    \includegraphics[width=\columnwidth]{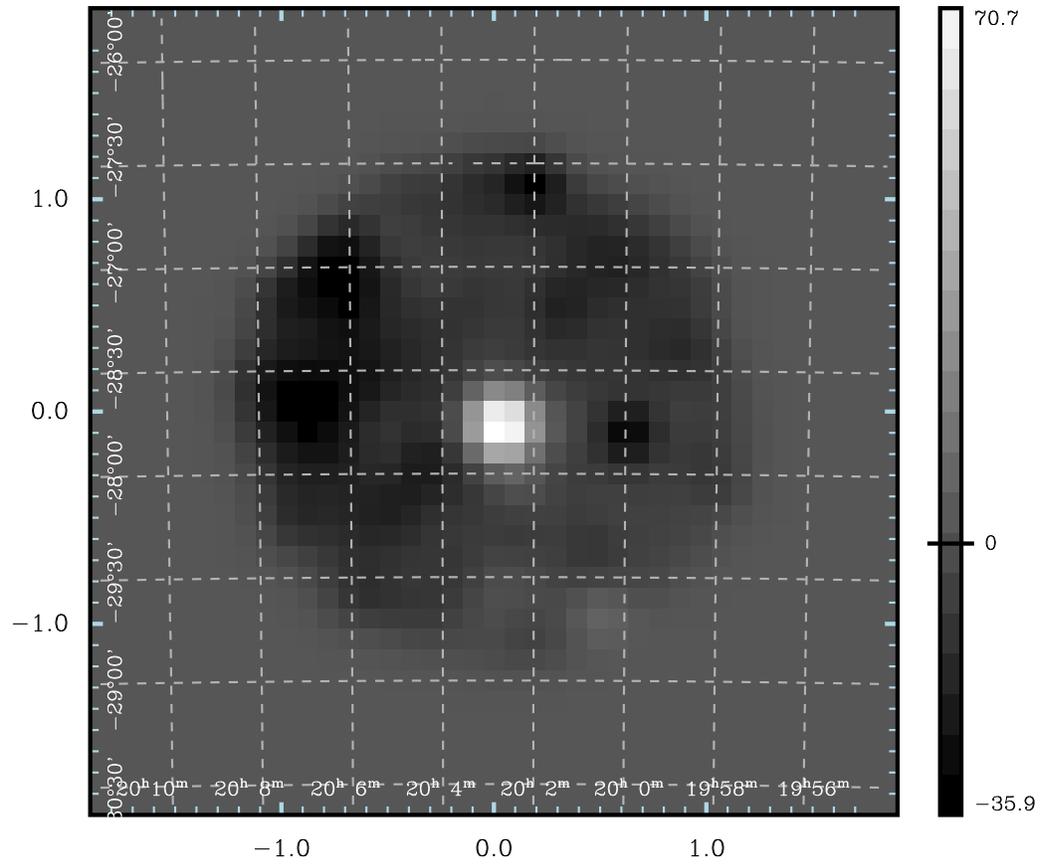}
    \caption{ \label{fig:offset} Optical image of a star
      ($\mathrm{Sgr}\ \gamma_2$) used to check the telescope's
      pointing at low elevation.  This image shows that the telescope
      has an offset of 0.14 degrees down and to the right of camera
      center position (0,0).}
  \end{figure}

  \begin{figure}[ht!]
    \includegraphics[width=\columnwidth]{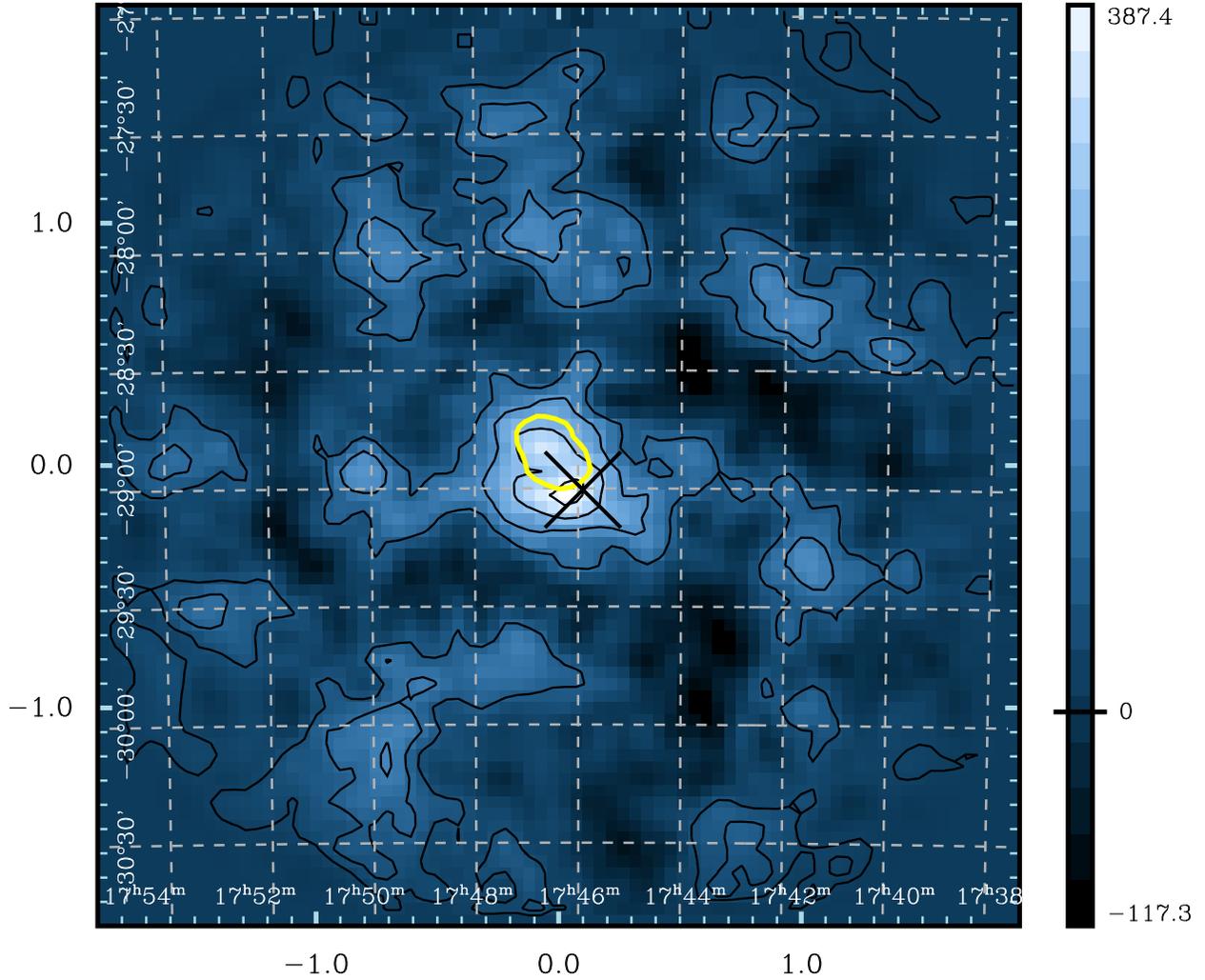}
    \caption{ \label{fig:sgra} A gamma-ray image of the region around
      Sgr A*. The image is of excess counts with overlaid significance
      contours (1 standard deviation per contour). The axes are labeled in
      degrees from the assumed camera center. The true center position
      of the camera, which is not exactly at (0,0) due to flexing of
      the telescope at low elevation, is marked with a cross.  The
      dashed lines are the RA and Dec contours at this position. Also
      shown (as a light contour) is the 99\% confidence region for the
      EGRET observations \citep{dingus2002}.}
  \end{figure}

  \begin{figure}[ht!]
    \includegraphics[width=\columnwidth]{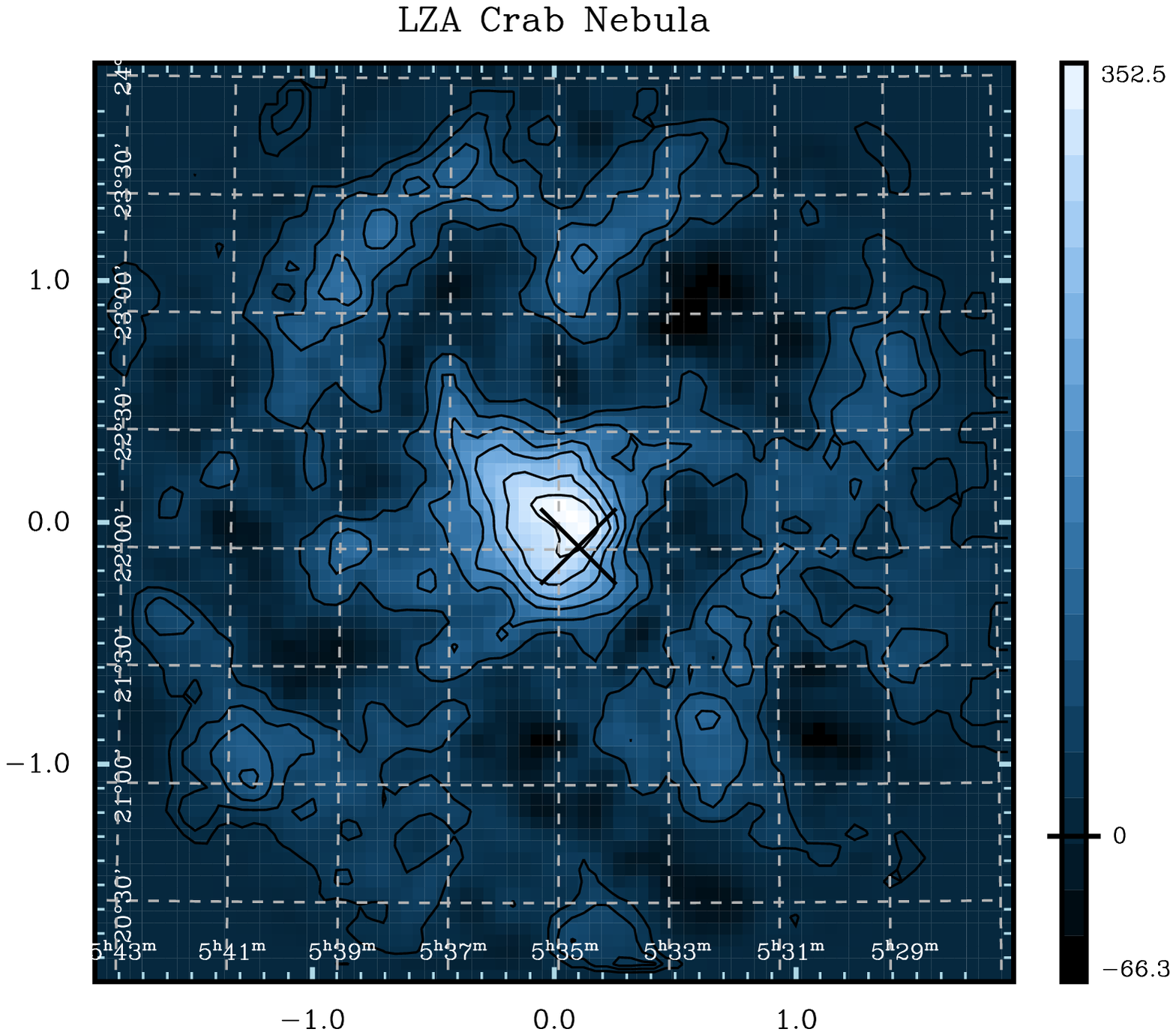}
    \caption{ \label{fig:lzacrab} A gamma-ray image of the Crab Nebula
      taken at large zenith angle ($\approx62^\circ$) using the same analysis
      procedure used for Sgr A*. The offset and pointing variations can be
      seen in the resulting image.}
  \end{figure}

  We have combined observations of Sgr A* from 1995 through 2003
  resulting in 26 hours of on-source exposure at an average zenith
  angle of $61^\circ$.  To determine the pointing offset, observations
  were taken centered on a nearby bright star ($\mathrm{Sgr}\
  \gamma_2$) which is at the same elevation as Sgr A*. Using the sky
  brightness map technique outlined earlier, we determined that the
  telescope had a pointing offset of $0.14^\circ$, as shown in Figure
  \ref{fig:offset}.  Figure \ref{fig:sgra} shows the resulting 2-D map
  of gamma-ray excess with overlaid significance contours. The true
  center of the camera, correcting for the offset, is plotted as a
  cross in the image.  This image shows a $4.2$ standard deviation
  ($\sigma$) excess at the corrected center position. To check the
  robustness of this result, we re-ran the analysis ten times to
  account for variations due to the Gaussian padding. We find the
  average significance at the corrected center position is $(3.7 \pm
  0.13)\ \sigma$, somewhat below the initial result. For reference, in
  Figure \ref{fig:lzacrab} we show the results of the same analysis
  procedure applied to 10 hours of observations of the Crab Nebula at
  a similar zenith angle range. Note that the significance of
  $6.1\sigma$ of the Crab detection at the offset position is
  substantially higher than the result of $3.8\sigma$ obtained
  applying the standard small zenith angle analysis procedure to these
  LZA data. Also, the similar angular extent in the two results
  indicates consistency with a point source within a 99\% confidence
  region of radius $\approx 15$ arcmin. A more detailed analysis of
  the angular extent would require a prior hypothesis about the nature
  of the source.

  

  To determine the peak energy of the detected flux from Sgr A*, we
  simulated gamma rays with a Crab Nebula spectrum (with integral
  spectral index $\gamma=1.58$)
  \citep{1998APh.....9...15M,2000ApJ...539..317A} and a zenith angle
  of $61^\circ$, and analyzed the resulting data with a detector
  simulation and our analysis software. We determined the peak
  detected energy to be $\approx 2.8\ \mathrm{TeV}$.  Noise padding
  and a higher trigger threshold makes this result slightly higher
  than has been previously reported \citep{1999ApJ...511..149K}, and
  we estimate a $20\%$ systematic error in this energy threshold.  We
  then analyzed a set of real LZA Crab Nebula data runs to find the
  Crab count rate and compared this to the corresponding rate for Sgr
  A*.  The integral flux for Sgr A*, normalized to the Crab, is then:
  \begin{equation}
    F_{SgrA}(>2.8\ \mathrm{TeV}) = N_{0,\mathrm{Crab}} \cdot
    \frac{(2.8\ \mathrm{TeV})^{-\gamma}}{\gamma} \cdot
    \frac{R_{\mathrm{SgrA*}}}{R_{\mathrm{Crab}}}
  \end{equation}
  Where $N_{0,Crab}$ is the flux normalization factor for the Crab
  Nebula ($3.12\times10^{-7} \mathrm{m^{-2}
  s^{-1}}$), $\gamma$
  is the integral Crab spectral index, and $R_{\mathrm{SgrA*}}$ and
  $R_{\mathrm{Crab}}$ are the corresponding Sgr A* and Crab Nebula
  gamma-ray count rates. From the LZA Crab data, we find a gamma-ray rate of
  $R_{\mathrm{Crab}}(>2.8\ \mathrm{TeV})=0.501 \pm
  0.087~\mathrm{photons\ min^{-1}}$ and from Sgr A* we obtain an
  average rate of $R_{\mathrm{SgrA*}}(>2.8\ \mathrm{TeV})=0.205 \pm
  0.057\ \mathrm{photons\ min^{-1}}$.  Hence, the gamma-ray
  flux from the Galactic Center region above 2.8 TeV is \Flux, or
  about 0.4 times that of the Crab Nebula (the flux error includes the
  uncertainty in the Crab Nebula measurement). 


  \begin{figure}[ht!]
    \includegraphics[angle=90,width=\columnwidth]{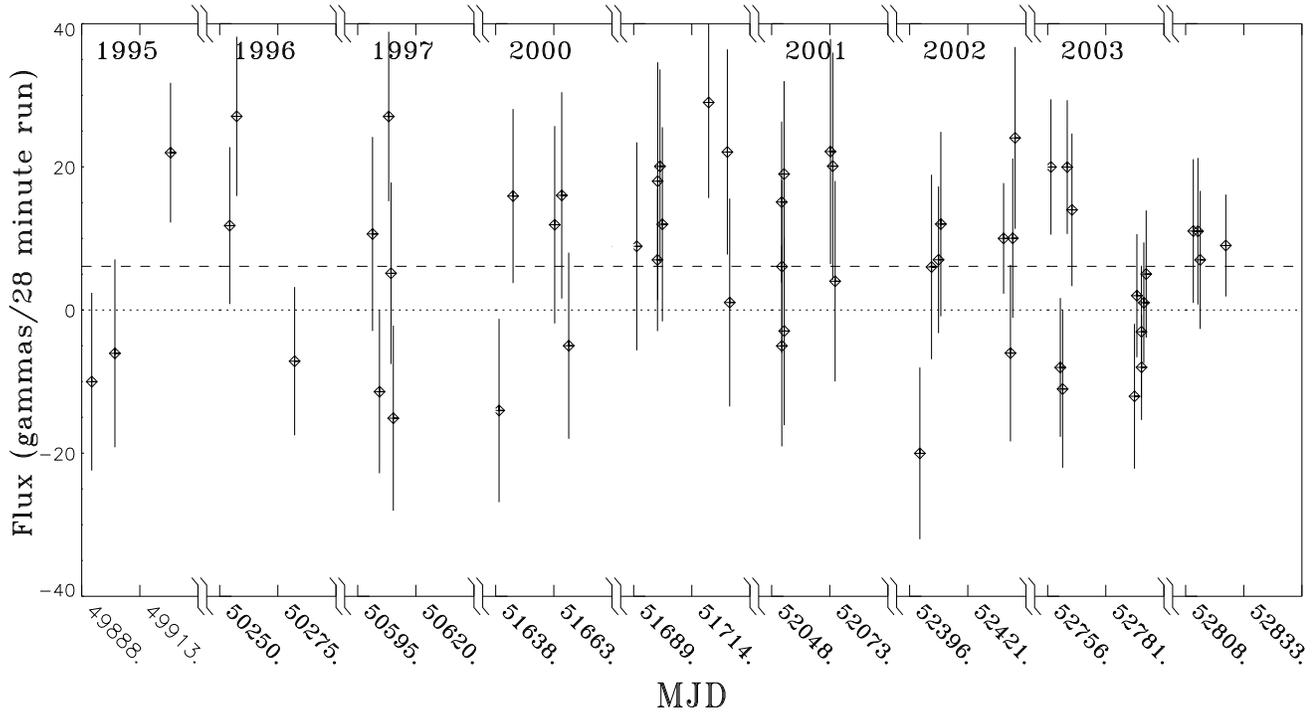}
    \caption{ \label{fig:gcvar} 
      Flux of Sgr A* as a function of time.  Each data point
      represents a single 28 minute run.  Time gaps in the data have
      been removed where indicated. The dashed line is a least squares fit
      of a constant function to the data. }
  \end{figure}

  The count rate from Sgr A* is shown as a function of time in
  Figure \ref{fig:gcvar}. To determine the probability for steady
  emission, a $\chi^2$ fit of a constant function ($f(t) = A$) was
  applied to this data and, for comparison, to a series of data taken
  of Mrk 421 (a source which is known to be highly variable) at a
  similar zenith angle range as Sgr A*. The total significance of this
  Mrk421 data sample was $2.3\sigma$. The Sgr A* data yields a
  constant count rate of $6.12 \pm 1.59 \ \gamma\cdot\mathrm{min}^{-1}$ with
  a reduced $\chi^2$ of $1.13$ (with 54 degrees of freedom), which
  corresponds to a $25\%$ probability that there is no
  variability. The result for Mrk~421 yields a constant count rate of $6.86
  \pm 6.13 \ \gamma\cdot\mathrm{min}^{-1}$ with a reduced $\chi^2$ of
  3.03 (with 6 degrees of freedom) and a $1.2\%$ chance of no
  variability.
  \section{Discussion}
  \label{sec:discussion}

  The TeV excess observed near the position of the Galactic Center is
  unlikely to have occurred by chance and constitutes a probable, as
  yet unconfirmed, detection of a new TeV source.  Possible
  systematics that could contribute to a false detection include the
  effects of additional noise from the relatively bright off-source
  region.  While we have largely corrected for these effects, some
  systematic uncertainties remain.  We have taken into account trials
  factors by formulating an explicit \emph{a priori} hypothesis that
  we would only look for emission at the exact position of the
  Galactic Center after a pointing correction was applied.
  Statistical variations in the analysis method (due to the addition
  of simulated noise in padding) have been taken into account by
  repeating the analysis ten times, and taking the average
  significance, giving a conservative estimate of $3.7\sigma$ for the
  detection significance.

  The lack of significant variability in our data makes it difficult
  to uniquely identify the source with a compact point source such as
  Sgr A*, but inspires some confidence in the stability of our
  observations at large zenith angle. Note that the analysis procedure
  was designed to mitigate against changes in the count rate due to
  variations in the instrument. The same ISU simulation package
  was used here to analyze Whipple observations of the Crab Nebula
  giving a spectrum in good agreement which that measured a small
  zenith angle \citep{1999ApJ...511..149K}.  In the past, our group
  reported a positive excess of $2.4\sigma$ for 1995-1997 observations
  \citep{buckley97} and $2.4\sigma$ for 1999-2003 observations
  \citep{kosack2003} at the position of Sgr A*.  The combined
  significance for our refined results is consistent with these
  earlier analyses.  The large error circles for both EGRET (7.2 arcmin) and
  Whipple (15 arcmin) observations make identification with a
  particular source difficult, but given the dearth of TeV sources, an
  accidental angular coincidence of a new source along the line of
  sight is unlikely, and it is probable that the emission comes from a
  non-thermal source physically near the Galactic Center.  

  The high level of emission $\approx 0.4\ \mathrm{Crab}$ at a
  distance of roughly four times that of the Crab Nebula, qualifies
  this as an unusually luminous galactic source.  Previous TeV
  observations of relatively nearby galactic sources such as X-ray
  binaries and shell-type and plerionic supernovae have produced
  numerous upper-limits, or (at best) unconfirmed detections, making
  the detection of such an object at 8.5kpc even more unlikely.
  \citet{1998A&A...335..161M} came to a similar conclusion about the
  GeV emission based on the high luminosity of the EGRET unidentified
  source and lack of significant variability. If the Sgr A East
  supernova shock were the source of the EGRET gamma rays, it would
  have been an unusually intense explosion \citep{1996ApJ...457L..61K}
  and a density of $1000\ \mathrm{cm^{-3}}$ and magnetic field of $B
  \sim 0.18 \mathrm{mG}$ (well above the canonical values) would be
  required \citep{2003ApJ...596.1035F}. While a typical galactic
  source such as an SNR, pulsar, or stellar mass black hole is
  unlikely, an active nucleus at our Galactic Center is still a viable
  possibility and the detection of correlated variability in future
  gamma-ray and X-ray observations is required to make the
  identification with Sgr A* compelling.  If we associate this
  emission with either the super-massive black hole Sgr A* or
  supernova remnant Sgr A East, the observed emission could come from
  self-Compton scattering by electrons with energies up to at least
  2.8 TeV or from pion-decay gamma-rays from primary protons of even
  higher energy (kinematics require their energy to be at least
  several times the maximum gamma-ray energy).

  The lack of significant variability and the consistency with the
  Galactic Center position allow more exotic possibilities such as the
  annihilation of very high mass ($> 2\ \mathrm{TeV}$) dark matter
  particles at the Galactic Center.  While not particularly
  constraining, these results and a more detailed analysis of the
  spectrum will be used to derive upper limits for dark matter
  annihilation in a subsequent paper.

  In summary, based on 26 hours of data taken with the Whipple 10m
  gamma-ray observatory, we report a probable detection of a gamma-ray
  point source consistent with the position of the Galactic Center.
  We also describe a modified analysis procedure that we have
  developed to analyze gamma-ray data taken by Imaging Atmospheric
  \Cerenkov telescopes operating at large zenith angles. The continued
  observation of TeV gamma rays from the Galactic Center has important
  theoretical implications for understanding a variety of
  astrophysical phenomena in the nuclear region of the Milky Way.

  \paragraph{Acknowledgements}
  {\small 
  We would like to thank Fluvio Melia, Paolo Gondolo, Johnathan Katz,
  and Ramanath Cowsik for useful discussion.
  The VERITAS Collaboration is supported by the U.S. Dept. of Energy,
  N.S.F., the Smithsonian Institution, P.P.A.R.C. (U.K.) and
  Enterprise-Ireland.}

\end{document}